# CHARACTERIZING THE GALAXY SHAPE EFFECT ON THE EUCLID SLITLESS SPECTROSCOPY USING SIMULATIONS

Louis Gabarra[1,2], on behalf of the Euclid Consortium

[1] Dipartimento di Fisica e Astronomia "Galileo Galilei", Università di Padova, 35131, Padua, Italy.
[2] INFN, Istituto Nazionale di Fisica Nucleare, Sezione di Padova, 35131 Padua, Italy

Correspondence: louis.gabarra@pd.infn.it

ABSTRACT

The next generation of wide spectroscopic surveys such as *Euclid* will scan the sky in the near-infrared to obtain both photometry and spectroscopy. For this purpose, the *Euclid* telescope will rely on a Near-Infrared Spectrometer and Photometer (NISP) instrument whose spectroscopic channel has been designed to operate in slitless configuration. This powerful and easy to operate configuration makes it possible to avoid any prior selection on the targeted galaxies while covering the entire field of view. Beyond the observed flux of the galaxy, the detector capabilities will strongly depend on the shape of the galaxy, which gets convolved with the galaxy spectra.
To test the effect of the galaxy shape on the quality of the *Euclid* slitless spectra, we have performed simulations testing potentially impactful morphological parameters. We then characterized the effect of the disk half-light radius on the quality of the *Euclid* slitless spectra.

INTRODUCTION

The *Euclid* mission (Laureijs et al., 2011; Racca et al., 2016) whose launch into space is expected for 2023, will scan the extragalactic sky in an under-sampled wavelength range, the near-infrared. This wavelength range provides access to the strong optical rest frame features, e.g., H$\alpha$ and [O$_{\text{III}}$]$\lambda$5008 emission, H$\beta$ absorption, up to redshift $\approx 2.5$, including the epoch where dark energy started to drive the accelerating expansion of the Universe. This redshift range is also crucial for galaxy evolution studies since it includes the cosmic noon (Madau & Dickinson, 2014), when galaxies were particularly prolific in terms of star-formation. The *Euclid* Wide survey (Euclid Collaboration: Scaramella et al., 2022) and its spectroscopic channel using the Near-Infrared Spectrometer and Photometer (NISP; Maciaszek et al., 2016) will therefore be of great use to infer physical parameters such as the star-formation rate, the dust attenuation, and the metallicity of galaxies in this redshift range. The physical parameters estimation inferred from spectroscopy will rely on *Euclid* ability to reconstruct emission lines. The *Euclid* Wide survey sensitivity, set at $2\times 10^{-16}\,\text{erg}\,\text{s}^{-1}\,\text{cm}^{-2}$ for a 0.25 arcsec radius source at 3.5$\sigma$, depends on the background level at the observed coordinates and on the shape of the galaxy. The latter is particularly important in slitless configuration. In this paper, we investigate the effect of the half-light radius of the galaxy on the *Euclid* capability to detect emission lines.

JUSTIFICATION

Slitless spectroscopy consists in removing any form of mask or slit, leaving the entire field of view (FoV) going through the spectrometer. *Euclid* slitless spectroscopy will be performed with a set of grisms. The grism is a dispersive and transmissive element in a collimated beam. Wide-field near-infrared slitless spectroscopy was first offered by the HST and now benefits from significant advances with previous studies on data reduction (Walsh et al., 1979; Pasquali et al., 2006; Ryan et al., 2018), and has therefore been retained to fulfill the *Euclid* requirements. Slitless spectroscopy benefits particularly from the space environment, getting rid of the telluric OH emission lines and avoiding the atmospheric absorption. Nevertheless, slitless spectra still present several challenges with respect to data reduction. Without any physical aperture restricting the spatial extension of the incoming light, the 2D spectra can be thought of as a convolution of the source and its spatial profile. This feature, while enabling spectroscopic study all over the FoV, is also responsible for potential contamination effects. There can be overlaps between different dispersion orders of different sources and self-contamination effects due to the finite extension of the observed galaxy, creating a spectral resolution dependance from the shape of the object.

In this work, we tested the effect of the galaxy shape on the quality of the *Euclid* slitless spectra. To identify the specific effects from the shape, we performed simulations where we ensured that spectra do not overlap, and we focused on the analysis on the first order spectra.

SIMULATIONS CONFIGURATION

*Frame and setup.* The simulation presented in this work have been done using the NISP spectroscopic pixel simulator (TIPS; Zoubian et al., 2014) that we have configured to mimic an *Euclid* Wide survey (Euclid Collaboration: Scaramella et al. 2022) spectroscopic acquisition, corresponding to an overall exposure time of 2304 s, resulting from four dithered frames of 576 s. These simulations include different sources of noise such as the read-out noise, the dark current, and the zodiacal light diffuse background. The latter is expected to be the dominant background and depends on the pointing coordinates. We chose pointing coordinates representative of a median zodiacal background level.

*Input preparation.* We produced an input catalog starting from one bright source of the COSMOS field (Laigle et al., 2016) that provided us with the physical parameters for this galaxy relying on multi-wavelength observations, from UV to far IR, to derive robust spectral energy distribution (SED) fitting parameters. The expected emission line fluxes for the wavelength range of interest were calculated using empirical calibrations available in the literature, i.e., star-formation rate-Hα relation, Baldwin-Phillips-Terlevich (BTP) diagram, mass-metallicity relation (MZR), and photoionization models (see Euclid Collaboration: Gabarra et al, in prep., for details). The resulting SED has then gone through TIPS together with a list of morphological parameters required to reproduce a realistic surface brightness distribution. We tested the effect of some parameters on the quality of the extracted spectra. These parameters are the orientation angle with respect to the dispersion axis, the inclination of the galaxy, the bulge fraction, and the disk half-light radius. In this paper, we report only the results of the simulations changing the disk half-light radius since it was found that this parameter is the one that most affects the quality of the spectra (see Euclid Collaboration: Gabarra et al, in prep.).

*Data Analysis.* We fitted the continuum with a linear interpolation of the extracted 1D spectra and subtracted it from the spectra. We assumed Gaussian profile for the emission lines and inferred from the fit the flux and full width at half maximum (FWHM). We fixed the central wavelength using the true redshift for each of the lines of interest and left the FWHM and amplitude as free parameters.

RESULTS

We present in Fig. 1.a an overview of the effect of the galaxy size on the extracted spectra by showing the results around the Hα line-complex, i.e., Hα blended with the [NII] doublet, obtained by emulating twice the same spectra but changing the Disk R50 from 0.5 to 1.5 arcsec.

To quantify the effect of the size on the spectral resolution, we present in Fig. 1.b the FWHM of the instrument ($FWHM_{inst}$) measured on the Hα line-complex as a function of the Disk R50. The $FWHM_{inst}$ is obtained by subtracting in quadrature the FWHM of the incident spectra (due to velocity dispersion) from the FWHM measured on the extracted spectra. This estimate accounted for the blending of the Hα line with the [NII] doublet due to the NISP spectral resolution. To this end we broadened the incident spectrum of an ideal point-like source with a Gaussian filter with FWHM = 25Å. We found that the FWHM of the Hα line-complex is about 1.3 times the FWHM of the single Hα line, and we applied this relation to all the sources. We followed the normalization presented in Pasquali et al. (2006) and normalized the $FWHM_{inst}$ by the *Euclid* NISP requirement for a 0.25 arcsec radius source ($\Delta\lambda_0$). The expected resolution ($R$) of the *Euclid* NISP-S, $R = \lambda/\Delta\lambda_0 > 380$ for a 0.25 arcsec radius source, is taken from Racca et al. (2016). The red line and circles are the median normalized $FWHM_{inst}$ calculated in Disk R50 bins including a fixed number of 50 sources. The error bars are the median absolute deviation (MAD). The solid black line is the result convolving the PSF inferred from the measurement on the NISP ground test campaign images (Costille et al., 2019; Maciaszek et al., 2022; Gillard et al., in prep), i.e., a 50% Encircling Energy radius (EE50) of 0.2 arcsec, with twice the projected disk half-light diameter. The factor two came from the fact that the fixed values for the position angle from the RGS simulator dispersion axis and for the inclination angle of the galaxy are both fixed at 45 degrees.

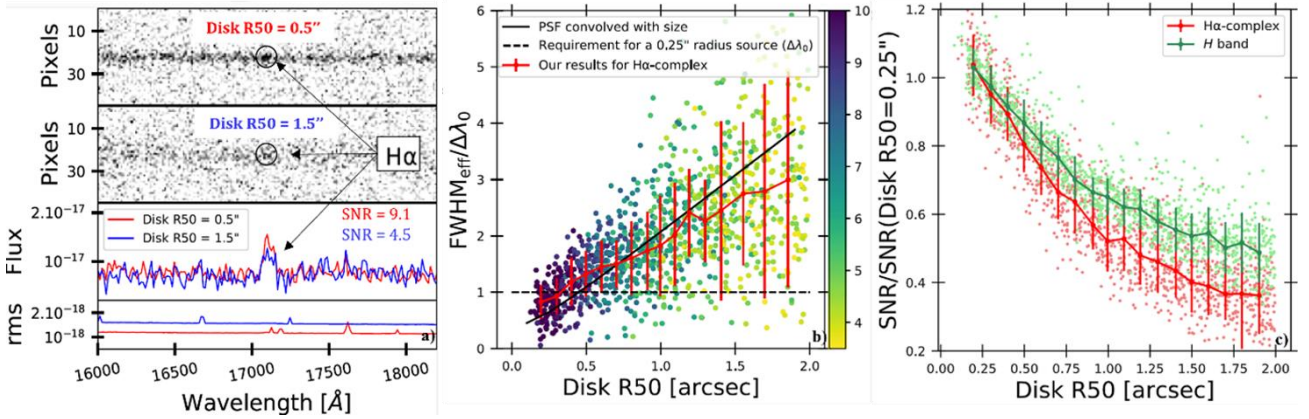

Figure 1. a) Illustration of the disk half-light radius effect on the quality of the spectra. The top two panels are the 2D simulated spectra obtained for the same input spectra with two different sizes, 0.5 and 1.5 arcsec. The bottom two panels are the flux and root mean square (rms). The red line indicates measurements on the 1D combined spectra 0.5 arcsec radius source while the blue line is for the 1.5 arcsec source. b) The spectral resolution measured on the extracted Hα line-complex as a function of the object Disk R50 in arcsec. The spectral resolution has been measured on a sample of 1248 sources with all morphological parameters set at a fixed value except for the Disk R50 that varies from 0.1 up to 2 arcsec. The red line shows the median normalized $FWHM_{inst}$ calculated on Disk R50 bins including a fixed number of 50 sources. The error bars show the median absolute deviation (MAD). The solid black line shows the model (see explanation in the text). c) SNR of the extracted Hα line-complex measurements (red) and extracted continuum measurements (green) normalized by the median SNR for sources with a disk half-light radius (R50) of 0.25 arcsec as a function of the Disk R50. The lines and error bars shows the normalized median SNR and MAD values calculated on Disk R50 bins including a fixed number of 50 sources. More details in Euclid Collaboration: Gabarra et al., in prep.

To evaluate the effect of the size on the quality of the measurement, we characterized the drop of the signal to noise ratio (SNR) as the Disk R50 increases. Results are shown in Fig. 1.c for the Hα line-complex flux measurement (in red) and for the continuum flux measurement in the H band (in green) that we normalized by the median SNR of a 0.25 arcsec Disk R50 source. The lines and circles show the median normalized SNR calculated on Disk R50 bins including a fixed number of 50 sources. The error bars show the MAD.

CONCLUSIONS

We showed that the disk half-light radius has a significant effect on the quality of the extracted slitless spectra and on our ability to detect emission lines. The $FWHM_{inst}$ scales as the Disk R50, degrading redshift determination accuracy and SNR. Beyond the almost linear degradation of the spectral resolution as the Disk R50 increases, we can observe in Fig. 1.b that the MAD significantly increases for sources with Disk R50 > 0.5 arcsec. We can also see that our measurements agree with the requirement set for a 0.25 arcsec radius source and with the expectation from the model. We characterized the effect of the disk half-light radius on the SNR and we showed that there is an almost linear drop of the SNR as the size increases, to reach a median degradation of ≈ 45 % for emission line and ≈ 35 % for continuum measurements for a source with Disk R50 equal to 0.1 arcsec compared to a 0.25 arcsec source.

*Acknowledgements*. The author thanks Chiara Mancini, Lucia Rodríguez-Muñoz, Giulia Rodighiero, Chiara Sirignano, Marco Scodeggio and Margherita Talia for their help in shaping this work. The Euclid Consortium acknowledges the European Space Agency and a number of agencies and institutes that have supported the development of *Euclid*, in particular the Academy of Finland, the Agenzia Spaziale Italiana, the Belgian Science Policy, the Canadian Euclid Consortium, the French Centre National d'Etudes Spatiales, the Deutsches Zentrum für Luft- und Raumfahrt, the Danish Space Research Institute, the Fundação para a Ciência e a Tecnologia, the Ministerio de Ciencia e Innovación, the National Aeronautics and Space Administration, the National Astronomical Observatory of Japan, the Netherlandse Onderzoekschool Voor Astronomie, the Norwegian Space Agency, the Romanian Space Agency, the State Secretariat for Education, Research and Innovation (SERI) at the Swiss Space Office (SSO), and the United Kingdom Space Agency. A complete and detailed list is available on the *Euclid* web site (http://www.euclid-ec.org).